\documentstyle[editedvolume]{crckapb} 
\def\gsim{\lower 2pt \hbox{$\, \buildrel {\scriptstyle >}\over
{\scriptstyle \sim}\,$}}
\def\lsim{\lower 2pt \hbox{$\, \buildrel {\scriptstyle <}\over
{\scriptstyle \sim}\,$}}
\def\rosat{{\sl ROSAT~}}
\def\asca{{\sl ASCA~}}

\begin{opening}
\title{X-ray Observations of the Hot Intergalactic Medium}\\

\author{Q. Daniel Wang}
\institute{Dearborn Observatory, Northwestern University\\
           2131 Sheridan Road, Evanston,~IL 60208-2900\\
		e-mail: wqd@nwu.edu}

\end{opening}

\runningtitle{The Hot Intergalactic Medium}

\begin{document}

\section{Introduction}

A definite prediction from recent N-body/hydro simulations of the structure 
formation of the universe is the presence of a 
diffuse hot intergalactic medium (HIGM; e.g., Ostriker \& Cen 1996).
The filamentary structure of the today's universe, as seen in various galaxies
surveys, is thought to be a result of the gravitational collapse of 
materials from a more-or-less uniform and isotropic early universe. During 
the collapse, shock-heating can naturally raise gas 
temperature to a range of $10^5-10^7$~K. Feedbacks from stars may also 
be an important heating source and may chemically enrich the HIGM. 
The understanding of the heating and chemical enrichment of the IGM 
is critical for studying the structure and evolution of clusters of
galaxies, which are nearly virialized systems (e.g., Kaiser 1991; 
David, Jones,\& Forman 1996). Most importantly, the HIGM may explain much 
of the missing baryon content required by the
Big Bang nucleosynthesis theories (e.g., Copi, Schramm, \& Turner 1995);
the total visible mass in galaxies and in the hot intracluster medium 
together is known to account for $\lsim 10\%$ of the baryon content 
(e.g., Persic \& Salucci 1992). 

	What are the probable observational signatures of the HIGM? 
First, the HIGM may contribute considerably to the soft X-ray background.  
Various emission lines of heavy elements (O, Ne, and Fe) may collectively form 
a distinct spectral bump at $\sim 0.7$~keV (Cen et al. 1995). This bump
has the potential as a diagnostic of the physical and chemical properties 
of the HIGM. Second, the density and temperature of the HIGM can be
greatly enhanced in filamentary superstructures near rich clusters
of galaxies because of deep gravitational potential. In these superstructures, 
X-ray emission from the HIGM can be greatly enhanced.

	Here I review lines of 
observational evidence for the HIGM. These results appear to be
consistent with the predictions from the simulations, and demonstrate
the potential of X-ray observations as a powerful tool to study the 
large-scale structure of the universe.

\section {Point-like versus Diffuse X-ray Components}

While $\gsim 75\%$ of the background in the 1 - 2 keV range is 
apparently due to point-like sources
(AGNs; Hasinger et al. 1993), the point-like contribution to the background 
decreases considerably at lower energies. 
Dick McCray and I have conducted an auto-correlation function (ACF) 
analysis of the X-ray background, based on \rosat observations (Wang \& McCray
1993). Using a multi-energy band ACF analysis technique, we 
characterized the mean spectrum of point-like sources, below a
source detection limit of $S(0.5-2 {\rm~keV}) \sim 1 \times 10^{-14} 
{\rm~ergs~s^{-2}~cm^{-2}}$, as a power law with $\alpha = 0.7 \pm 0.2$. 
This spectrum is significantly flatter than that of the 
observed background. We find that 
point-like sources cannot account for more than about $ 60\%$ 
of the background in the M-band ($\sim 0.4-1$~keV) without exceeding 
the total observed flux 
in higher energy bands. To explain the background
spectrum, an additional thermal component of a characteristic
temperature of $\sim 2 \times 10^6$~K is required, in addition to the 
well-constrained local $10^6$~K component. 

\section {Spectroscopic Evidence for a Thermal Component}

	Direct evidence for a {\sl thermal} M-band component of the 
background comes from the {\sl ASCA} X-ray Observatory, which provides 
spectroscopic capability between 0.4-10~keV. The \asca spectrum
of the X-ray background clearly shows an excess in the M-band range
above the extrapolation of the power law ($\alpha \approx 0.4$) that
fits well to the 1-10~keV spectrum. The excess also shows signs of OVII and 
OVIII lines, suggesting that at least part of it is thermal 
in origin (Gendreau et al. 1995). There are, however, some significant
discrepancies (up to $\sim 30\%$) between soft X-ray background normalizations 
derived from different instruments (McCammon \& Sanders 1990; Wu et al. 1991;
Garmire et al. 1992; Gendreau et al. 1995; Snowden et al. 1995),
making it hard to compare different measurements of absolute background 
intensities. Nevertheless, both the \rosat ACF analysis 
and the \asca spectrum do seem to 
suggest that a considerable fraction of the M-band background
arises in diffuse thermal gas. The question is where the gas is located:
Galactic or extragalactic?

\section {Galactic Foreground versus Extragalactic Background}

	While the extragalactic nature of the hard X-ray background 
in the 1 - 100 keV range is now generally accepted, 
a considerable Galactic contribution is expected at lower energies. 
X-ray shadowing studies are probably the
only viable approach to separate the Galactic foreground from
the extragalactic background. Such studies have yielded various estimates of
the extragalactic background intensity in the range of 
26 - 62 ${\rm~keV~s^{-1}~cm^{-2}~keV^{-1}} {\rm~sr^{-1}}$ at $\sim 0.25$~keV
(e.g., Snowden et al. 1994; Cui et al. 1996; Barber, Roberts, 
\& Warwick 1996). The uncertainty in these estimates is still
too large, however, to place a useful constraint on the diffuse HIGM 
contribution.

	We have recently measured the extragalactic background at 
$\sim 0.7$~keV by observing the X-ray shadowing of a neutral gas cloud in the 
Magellanic Bridge region (Wang \& Ye 1996). The cloud is at a distance of
$\sim 60$~kpc and has a peak HI column density of a few times $10^{21}
{\rm~cm^{-2}}$. From the anti-correlation 
between the observed background intensity and the HI column density of the 
cloud, we derived an unabsorbed extragalactic background intensity as $\sim 28
{\rm~keV~s^{-1}~cm^{-2}~keV^{-1}} {\rm~sr^{-1}}$
at $\sim 0.7$~keV, with the 95\% confidence lower limit as $18
{\rm~keV~s^{-1}~cm^{-2}~keV^{-1}} {\rm~sr^{-1}}$. Part of this extragalactic
emission must come from point-like sources such as AGNs. But the average
spectrum of point-like  sources is significantly flatter than that of 
the background and apparently flattens with decreasing source fluxes, as
shown in our ACF analysis. Using the total observed background intensity
in the 1-2~keV band, we obtain an upper limit to the source contribution 
as $\lsim 14 
{\rm~keV~s^{-1}~cm^{-2}~keV^{-1}} {\rm~sr^{-1}}$, which is smaller than
the 95\% confidence lower limit of the measured extragalactic background
(i.e., $18-14=4 {\rm~keV~s^{-1}~cm^{-2}~keV^{-1}} {\rm~sr^{-1}}$). 
Although the actual intensity of the diffuse component is 
still greatly uncertain, our measurement is consistent with 
the HIGM contribution to the M-band background as predicted with popular
cosmological models (Cen et al. 1995). 

\section {HIGM Features near Rich Clusters}

	As illustrated in the simulations, one may also expect to detect
enhanced X-ray-emitting filamentary superstructures near rich clusters of 
galaxies. In a pilot study of the environs of 
intermediate redshift clusters, A. Connolly, R. Brunner, and I (1997)
have discovered that A2125 is within a filamentary
complex consisting of various extended X-ray-emitting features. We have 
further made a multi-color optical survey of galaxies 
in the field, using the Kitt Peak 4-m telescope. The color distribution
of galaxies in the field 
suggests that this complex represents a hierarchical superstructure
spanning $\sim 11h_{50}^{-1}$~Mpc at redshift $\sim 0.247$.  The
multi-peak X-ray morphology of A2125 suggests that the cluster is an
ongoing coalescence of at least three major subunits.  The dynamic
youth of this cluster is consistent with its large fraction of blue galaxies
observed by Butcher \& Oemler (1984).  The 
complex also contains two additional clusters. 
But the most interesting feature is the large-scale low-surface-brightness 
X-ray emission from a moderate galaxy concentration associated with
the complex. 

	This hierarchical complex, morphologically and spectrally
very similar to superstructures seen in various
N-body/hydrodynamic simulations (e.g., CDM$+\Lambda$ universe; Cen \&
Ostriker 1994). We find that such superstructures can naturally explain the positive
cross-correlation function between nearby Abell clusters and the 
X-ray background
surface brightness at $\sim 1$~keV, as detected by Soltan et
al. (1996). 

	While these results have offered us a first glimpse of the
HIGM, observations with upcoming X-ray observatories such as AXAF, XMM,
and Astro-E, will enable us to greatly improve the measurements. Detailed
comparisons with the simulations will become possible, providing 
unique information on various physical and chemical processes of the 
structure formation of the universe. 

\section {References}
%\begin{quote}
{\small
\begin{verbatim}
Barber, C. R., Roberts, T. P., & Warwick, R. C. 1996, MNRAS, 282, 157
Butcher, H., & Oemler, Jr. A. 1984, ApJ, 285, 426
Cen, R., & Ostriker, J. P. 1994, ApJ, 429, 4
Cen, R., Kang, H., Ostriker, J. P., & Ryu, D. 1995, ApJ, 451, 43
Copi, C. J., Schramm, D. N., & Turner, M. S. 1995, Science, 267, 192
Cui, W., et al. 1996, ApJ, 468, 117
David, L. P., Jones, C., & Forman, W. 1996, ApJ, 473, 692
Garmire, G. P., et al. 1992, ApJ, 399, 694
Gendreau, K. C., et al. 1995, PASJ, 47, L5
Hasinger, G., et al. 1993, A&A, 275, 1
Kaiser, N. 1995, ApJ, 383, 104
McCammon, D., & Sanders, W. T. 1990, ARA&A, 28, 657      
Ostriker, J. P., & Cen, R. 1996, ApJ, 464, 27
Persic, M., & Salucci, P. 1992, MNRAS, 258, 14p
Snowden, S. L., et al., 1994, ApJ, 430, 601
Snowden, S. L., et al., 1995, ApJ, 454, 643
Soltan, A. M., et al. 1996, A&A, 305, 17
Wang, Q. D., Connelly, A., & Brunner, R. 1997, ApJL, 487, 13
Wang, Q. D., & McCray, R. 1993, ApJL, 409, 37
Wang, Q. D., & Ye, T. 1996, New Astronomy, 1, 245
Wu, X.-Y., Hamilton, T. T., Helfand, D. J., & Wang, Q. D. 1991, ApJ, 379, 564
\end{verbatim}}
%\end{quote}

\end{document}